\documentclass[aip, apl, superscriptaddress, longbibliography, reprint]{revtex4-2}

\usepackage{graphicx}
\usepackage{bm}
\usepackage{dcolumn}
\usepackage{amssymb,amsmath}
\usepackage[colorlinks=true,allcolors=blue]{hyperref}
\usepackage{color}
\usepackage{siunitx}
\usepackage{placeins}

\usepackage{lipsum}
\newcommand{\cOut}[1]{}
\newcommand{\figref}[2]{\hyperref[#1]{\ref{#1}(#2)}}

\begin{document}

\title{Strain-induced frequency pulling in CoFeB/Cu/Py double-vortex oscillators}

\author{Vadym Iurchuk}
\email[Corresponding author's e-mail: ]{v.iurchuk@hzdr.de}
\affiliation{Institute of Ion Beam Physics and Materials Research, Helmholtz-Zentrum Dresden-Rossendorf, 01328 Dresden, Germany}

\author{J\"urgen Lindner}
\affiliation{Institute of Ion Beam Physics and Materials Research, Helmholtz-Zentrum Dresden-Rossendorf, 01328 Dresden, Germany}

\author{J\"urgen Fassbender}
\affiliation{Institute of Ion Beam Physics and Materials Research, Helmholtz-Zentrum Dresden-Rossendorf, 01328 Dresden, Germany}
\affiliation{Institute of Solid State and Materials Physics, Technische Universit\"at Dresden, 01062 Dresden, Germany}

\author{Attila K\'akay}
\affiliation{Institute of Ion Beam Physics and Materials Research, Helmholtz-Zentrum Dresden-Rossendorf, 01328 Dresden, Germany}

\date{\today}

\begin{abstract}
We demonstrate piezostrain-induced frequency pulling in stacked double-vortex structures, magnetostatically coupled through the nonmagnetic spacer. We study the effect of the Cu spacer thickness on the strain-induced gyrotropic frequency shift in double-vortex structures comprising of magnetostrictive CoFeB and nonmagnetostricitve Py layers. For the two stacked vortices with different eigen-frequencies, the strain-induced magnetoelastic anisotropy leads to the downshift of the gyration frequency of the magnetostricitve vortex. We show that for increased dipolar coupling between the layers (\textit{i.e.} decreased spacer thickness), a strain-induced frequency pulling regime is obtained, where the resonance frequency of the nonmagnetostrictive Py vortex is upshifted towards the gyration resonance of the magnetostrictive CoFeB vortex. This result offers an additional degree of freedom for the manipulation of the dynamical regimes and synchronization conditions in spintronic oscillators, controlled by voltage and tunable by strain
\end{abstract}

\maketitle

Nanoscale circular ferromagnetic disks with specific geometric aspect ratios naturally adopt a stable vortex configuration with in-plane magnetization circulating around a vortex core pointing out-of-plane~\cite{aharoni_upper_1990, usov_magnetization_1993, shinjo_magnetic_2000}. Resonant excitation of the vortex core by an external driving force at the specific frequency drives its so-called gyrotropic motion at a distinct eigen-frequency (usually in sub-GHz range), defined mainly by the ferromagnetic material parameters and the disk size~\cite{guslienkoEigenfrequenciesVortexState2002}.
Implementation of magnetic vortices as oscillating layers in spin-torque oscillators enables the generation of rf signals in nW amplitude range with high quality factors~\cite{dussauxLargeMicrowaveGeneration2010a}. Recent advancements have demonstrated that vortex-based oscillators are suitable candidates for emerging analog spintronic devices showcasing the potential for ultrafast spectrum analysis~\cite{litvinenkoUltrafastSweepTunedSpectrum2020b}, wireless communication\cite{litvinenkoAnalogDigitalPhase2019} and neuromorphic computing~\cite{romeraVowelRecognitionFour2018}. Despite their advantageous features as nanoscale rf elements, a notable limitation of vortex-based nano-oscillators is their limited frequency tunability when operated in the linear regime. A promising approach to overcome this limitation is to use strain-induced magnetoelastic anisotropy for the broadband tuning of the vortex gyrotropic frequency~\cite{iurchukPiezostrainLocalHandle2023}. Moreover, recent studies of the strain-tunable control over the magnetization reversal in spin-valve-like structures comprising magnetostrictive and nonmagnetostrictive layers demonstrated that magnetostrictively optimized multilayers offer a path to enhanced strain-controllable spintronic devices~\cite{iurchukStraincontrolledMagnetostrictivePseudo2023}.

The reported here experimental study is inspired by a micromagnetic investigation of the gyration dynamics of stacked double-vortex structures under external mechanical stress~\cite{iurchukStressinducedModificationGyration2021}. We study the effect of the Cu spacer thickness on the strain-induced gyrotropic frequency shift in double-vortex structures comprising magnetostrictive (Co$_{40}$Fe$_{40}$B$_{20}$, hereafter CoFeB) and nonmagnetostricitve (Py) layers. For the two stacked vortices with different eigen-frequencies, a strain-induced magnetoelastic anisotropy leads to a downshift of the gyration frequency of the magnetostricitve vortex. We show that for increased dipolar coupling between the layers (i.e. decreased spacer thickness), the strain-induced frequency pulling regime can be achieved, where the resonance frequency of the nonmagnetostrictive Py vortex is upshifted towards the gyration resonance of the magnetostrictive CoFeB vortex. This result demonstrates the strain-controlled tuning of the gyration dynamics in stacked magnetic vortices, and can be applied to the synchronization of the vortex-based spintronic oscillators. 

\begin{figure}[b]
\centering
    \includegraphics[width=0.5\textwidth]{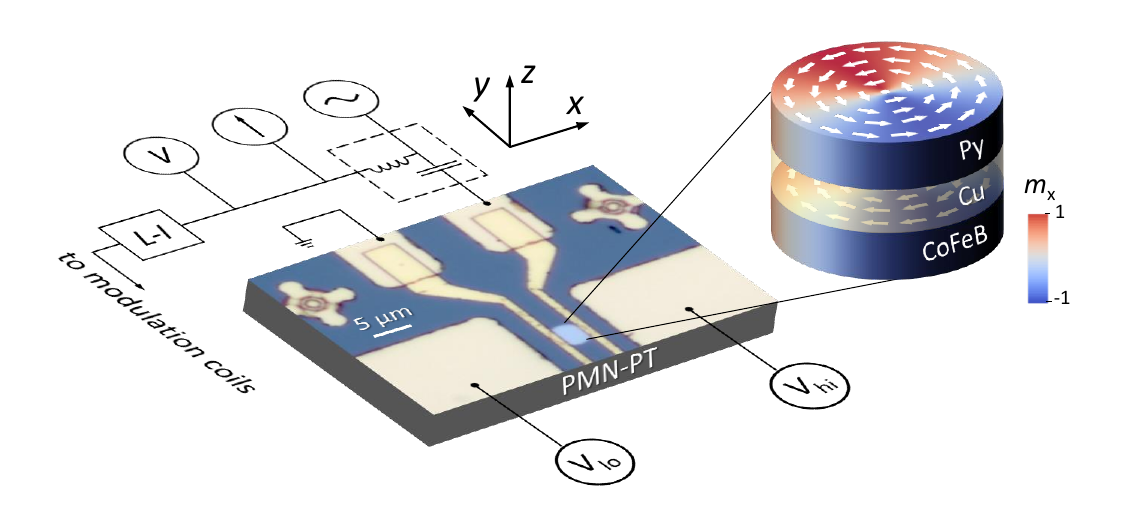}
    \caption{Schematics of the experimental setup for the magnetization dynamics detection in magnetic microdisks by spin rectification measurements, with a simultaneous application of an electric field to the piezoelectric PMN-PT substrate. The inset shows the double-vortex structure under study, comprising magnetostrictive CoFeB, nonmagnetic Cu and nonmagnetostrictive Py layers.} 
    \label{fig1}
\end{figure}

\begin{figure}[t]
\centering
    \includegraphics[width=0.5\textwidth]{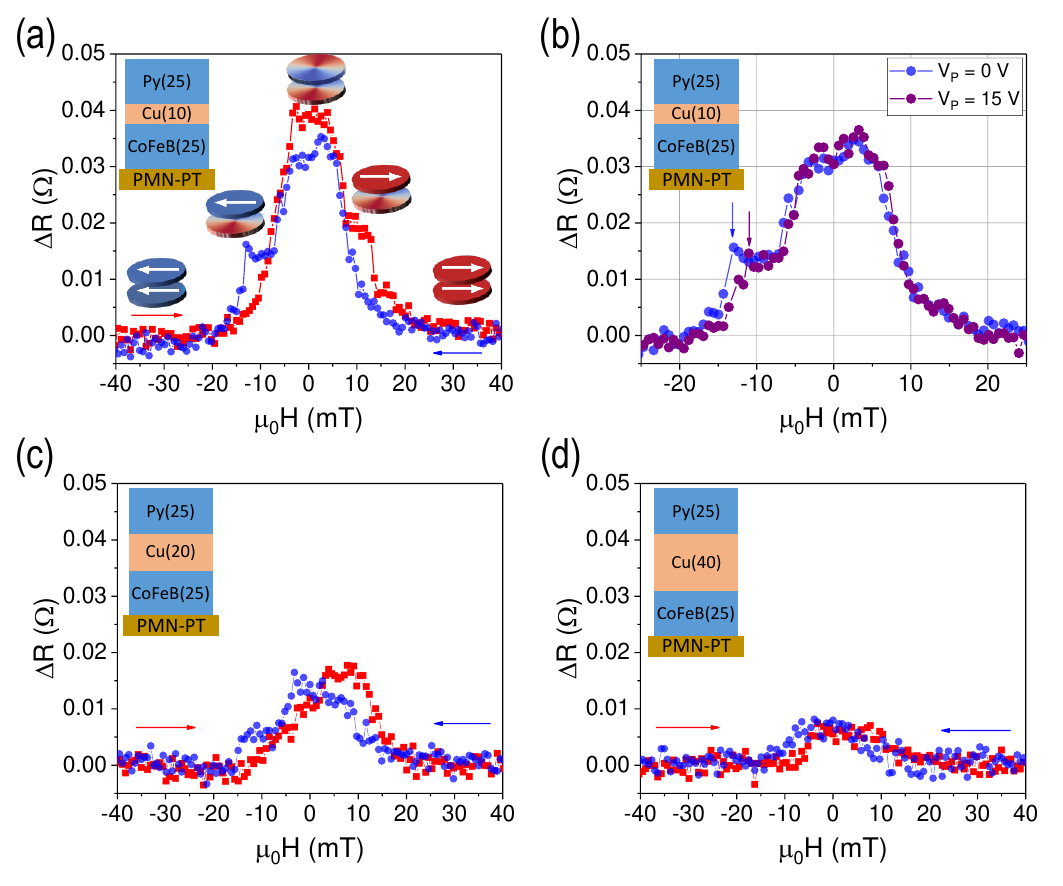}
    \caption{(a) Magnetoresistance of the CoFeB(25)/Cu(10)/Py(25) double-disk structure measured at $I_{dc}$ = 2~mA for the field applied in-plane and perpendicular to the probing current direction. The arrows indicate the field sweep direction. The insets sketch the magnetization configurations of the CoFeB and Py layers within the corresponding field ranges. (b) Magnetoresistance of the CoFeB(25)/Cu(10)/Py(25), measured at $V_P$ = 0 (blue circles) and $V_P$ = 15~V (purple triangles). (c,d) Magnetoresistance of the CoFeB(25)/Cu(20)/Py(25) (c) and CoFeB(25)/Cu(40)/Py(25) (d) double-disk structures measured for the same conditions as (a).} 
    \label{fig2}
\end{figure}

Fig.~\ref{fig1} shows the photomicrograph of the sample with schematic representation of the CoFeB/Cu/Py double-vortex structure and the experimental setup used for the electrical detection of the magnetization dynamics. We use (011)-cut 0.7Pb[Mg$_{1/3}$Nb$_{2/3}$]O$_3$--0.3PbTiO$_3$ (hereafter PMN-PT) single crystals as piezoelectric substrates to generate piezostrain upon application of a static electric field. Surface electrodes, magnetic double-disk stacks and contact pads were fabricated on the PMN-PT substrates in a three-step lithography process, similar to that described in~\textcite{iurchukPiezostrainLocalHandle2023}.
The surface electrodes to generate a local strain in the PMN-PT substrate, were fabricated by UV lithography, e-beam metallization with Cr(5~nm)/Au(125~nm) and conventional lift-off. The double-vortex stacks were patterned as 4~$\mu$m disks by means of electron beam lithography, followed by magnetron sputtering of a Cr(3)/CoFeB(25)/Cu($t_{Cu}$)/Py(25)/Cr(2) films (the thicknesses in parentheses are given in nm) and a lift-off. We fabricated a set of three samples with the Cu thickness $t_{Cu}$ = 10, 20 and 40~nm.
The bottom and top Cr layers were used as seed and cap layers, respectively. As the final step, contact pads were fabricated by electron beam lithography, e-beam evaporation of Cr(5~nm)/Au(50~nm) and lift-off to provide individual electrical access to each microdevice.

To detect the vortex core dynamics, we used a standard room-temperature magnetotransport setup with rf capability (Fig.~\ref{fig1}) for electrical detection of the rf-current-driven magnetization dynamics via rectification of the dynamic anisotropic magnetoresistance signal at the excitation frequency. More details about the detection technique can be found in~\cite{ramasubramanian_effects_2022, ramasubramanian_thesis_2022, iurchukPiezostrainLocalHandle2023}. We measured the vortex core dynamics in CoFeB/Cu/Py stacks vs. a dc voltage $V_P = V_{hi} - V_{lo}$ applied between the surface electrodes (Fig.~\ref{fig1}), yielding an electric field of 0.1 MV/m per unit voltage $V_P$.

Fig.~\ref{fig2}(a) shows a typical magnetoresistance of the CoFeB(25)/Cu(10)/Py(25) disk with 4~$\mu$m diameter measured for the dc current $I_{dc}$ = 2~mA applied through the contact pads. The magnetic field is applied in-plane and perpendicular to the direction of the dc current flow. Upon increasing the field from the negative saturation [red curve in Fig.~\ref{fig2}(a)], a slight resistance increase is observed at about --15~mT, attributed to the onset of the vortex nucleation at the edges of the ferromagnetic disks through the formation of buckling states. At approximately --10~mT, the resistance raises substantially indicating the vortex formation inside the disks. Due to strong dipolar coupling the vortex nucleation fields for CoFeB and Py disks are very close, which leads to  almost simultaneous vortex formation in both disks at a given external field value. 

When the field approaches zero, we observe a gradual increase of the resistance as both CoFeB and Py vortex cores propagate towards the center of the respective disk. Further field increase above zero leads to two notable resistance drops at $\sim$10~mT and $\sim$15~mT. The first drop is attributed to the vortex expulsion from the Py disk, whereas the second corresponds to the vortex expulsion for the CoFeB disk. Eventually, above 20~mT both disks are in saturated state.
When the field is swept from the positive saturation, a symmetric picture is observed. A slight asymmetry in the switching fields and the resistance values for the opposite field sweep directions may be attributed to the presence of a small out-of-plane field component due to the field inhomogeneities between the magnet poles.

We assume and also shown by simulations that, in the vortex pair, in the vicinity of $H$=0, the magnetization circulation directions within Py and CoFeB vortices are opposite to each other, since this configuration is energetically more favourable as compared to the parallel circulations. Dynamic measurements, as discussed further, suggest that the core polarities are antiparallel. 

\begin{figure}[t]
\centering
    \includegraphics[width=0.5\textwidth]{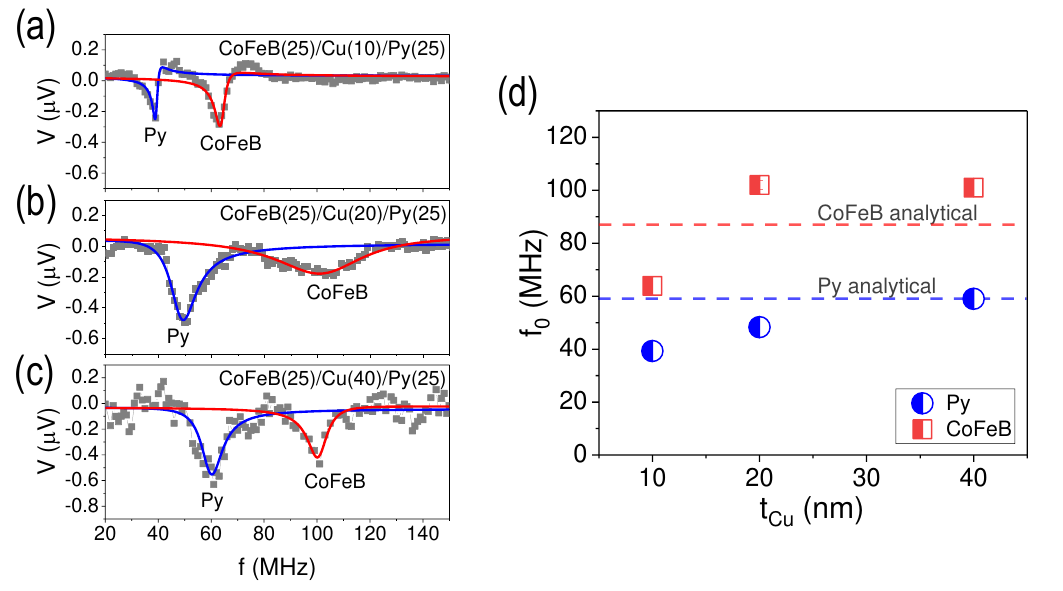}
    \caption{(a--c) Rectification spectra of CoFeB(25)/Cu($t_{Cu}$)/Py(25) double vortex structures for $t_{Cu}$ = 10~nm (a), 20~nm (b) and 40~nm (c). Solid lines are fits to the sum of symmetric and antisymmetric Lorentzian functions. (d) Resonance frequencies extracted from fitting the experimental data in (a--c). Dashed lines correspond to the analytically calculated gyrotropic frequencies of isolated CoFeB(25) and Py(25) disks with 4~$\mu$m diameter.} 
    \label{fig3}
\end{figure}

Fig.~\ref{fig2}(b) shows the magnetoresistance of the CoFeB(25)/Cu(10)/Py(25) disk measured for $V_P$=0 and 15~V applied to the PMN-PT via surface electrodes. One can see that the CoFeB vortex core expulsion field decreases from 15 to 12~mT with applied voltage, as expected due to the strain-induced magnetoelastic anisotropy (see~\cite{iurchukPiezostrainLocalHandle2023} for details]). On the other hand, we did not observe any changes to the nucleation field. This can be explained by the strong dipolar coupling of the magnetostrictive CoFeB vortex to the nonmagnetostrictive Py, leading to the nucleation of both vortices at the same field value.
On the contrary, when the field approaches the repulsion field for CoFeB vortex, the Py disk is already almost saturated (or rather is in a C-state) providing an additional dipolar field acting on the CoFeB layer. At this configuration, the magnetoelastic anisotropy facilitates the field-induced vortex core expulsion.

Figs.~\ref{fig2}(c,d) show the magnetoresistance of CoFeB(25)/Cu(20)/Py(25) and CoFeB(25)/Cu(40)/Py(25) disks. We observe progressively weaker MR signal for increased thicknesses of the Cu spacer. An origin of the reduced MR ratio is the increased fraction of the shunt current through the thick Cu layers and, therefore, reduced effective probing current flowing through the magnetic layers~\cite{10.1063/1.3337739}. Already for $t_{Cu}$ = 40~nm, the switching fields become indistinguishable [see Fig.~\ref{fig2}(d)].
For the devices with $t_{Cu}$ = 20 and 40~nm, magnetoresistance measurements for different $V_P$ applied to PMN-PT did not allow to detect any effects of strain on the magnetization switching in CoFeB and Py disks.

Figs.~\ref{fig3}(a--c) show rectification spectra of CoFeB(25)/Cu($t_{Cu}$)/Py(25) double vortex structures for $t_{Cu}$ = 10, 20 and 40~nm respectively. The $V_{dc}$ spectra in Fig.~\ref{fig3}(a) were measured at zero dc current, zero bias field and for --3~dBm of injected power from the rf generator. Larger rf power of --2~dBm and 6~dBm were used to measure the spectra in Figs.~\ref{fig3}(b,c) respectively. In addition, a small bias field (approximately 2~mT for the measurement in Fig.~\ref{fig3}(b) and 4~mT for Fig.~\ref{fig3}(c)) was applied in the $Oy$ direction to obtain better signal-to-noise ratio.

For all three thicknesses of the Cu spacer, we observe two resonances, which are attributed to the Py-dominated (lower frequency $f_{Py}$) and CoFeB-dominated (higher frequency $f_{CoFeB}$) gyrotropic dynamics in the coupled two-vortex structure. 
The fitting of the experimental data with a sum of symmetric and antisymmetric Lorentzian functions allows for the extraction of the corresponding gyrotropic frequencies, which are plotted in Fig.~\ref{fig3}(d) versus the spacer thickness $t_{Cu}$. 
When $t_{Cu}$ increases, both $f_{Py}$ and $f_{CoFeB}$ increase and approach their limit values defined by the analytically calculated gyrotropic frequencies of the noninteracting CoFeB and Py vortices. On the other hand, for the smallest $t_{Cu}$=10~nm, not only $f_{Py}$ and $f_{CoFeB}$ are significantly smaller as compared to the analytical limit, but the difference $f_{CoFeB} - f_{Py}$ decreases too. This result is in agreement with the micromagnetic study of the impact of the dipolar coupling on the gyration dynamics in stacked vortices~\cite{iurchukStressinducedModificationGyration2021}.
In addition, the observed spacer thickness dependence is the signature of the antiparallel mutual orientation of the CoFeB and Py vortex cores, as shown in~\cite{iurchukStressinducedModificationGyration2021}.

\begin{figure}[b]
\centering
    \includegraphics[width=0.5\textwidth]{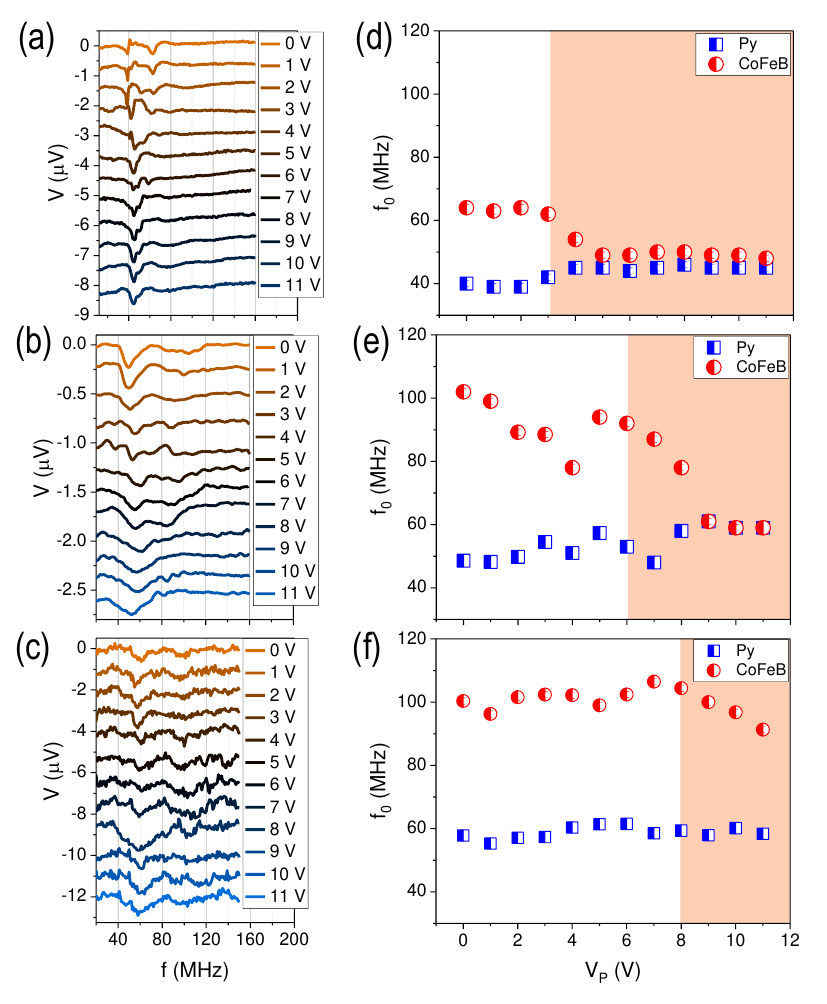}
    \caption{(a--c) Rectification spectra of CoFeB(25)/Cu($t_{Cu}$)/Py(25) double vortex structures for $t_{Cu}$ = 10~nm (a), 20~nm (b) and 40~nm (c), measured for different values of the voltage $V_P$ applied to the PMN-PT. (d--f) Resonance frequencies extracted from fitting the experimental data in (a--c) plotted as a function of the voltage $V_P$. Squares denote the Py-dominated resonance, circles denote the CoFeB-dominated resonance.} 
    \label{fig4}
\end{figure}

Figs.\ref{fig4}(a-c) show the spectra of  double vortex structures for $t_{Cu}$ = 10~nm (a), 20~nm (b) and 40~nm (c) measured as a function of the voltage $V_P$ applied to the PMN-PT substrate.
Fig.~\ref{fig4}(d--f) show the resonant frequencies extracted from the Lorentzian fits of the experimental spectra of Figs.~\ref{fig4}(a--c). Here, squares denote the Py-dominated gyrotropic resonance and circles correspond to the CoFeB-dominated resonance.
Similar to Fig.~\ref{fig3}(a), we observe an increased noise in the rectification spectra for $t_{Cu}$ = 20 and 40~nm as compared to the data for $t_{Cu}$ = 10~nm. The origin of the reduced signal amplitude as compared to noise is the gradually decreased value of the probing current flowing through the magnetic layers as the result of the increased shunt current into the progressively thicker Cu layers, as discussed above. The associated drop in the magnetoresistance also contributes to the reduced rectification voltage amplitude.

As seen from Fig.~\ref{fig4}(d), the resonance frequency, associated with the gyrotropic dynamics in the CoFeB disk (plotted with red circles) decreases with increasing $V_P$. This effect is the expected consequence of the strain-induced magnetoelastic anisotropy, which softens the vortex core restoring force spring constants~\cite{roy_2013}, leading to the reduced vortex core gyrotropic frequency. On the other hand, one would expect that the magnetoeastic anisotropy introduces only negligible effects on the vortex core dynamics in nonmagnetostrictive Py disks. However, as observed in the dynamical experiments on the CoFeB(25)/Cu(10)/Py(25) double-vortex device [Fig.~\ref{fig4}(d)], the Py vortex gyrotropic frequency is upshifted as the voltage $V_P$ (and so the strain acting on the magnetic structure) increases. This increase of the Py gyrofrequency occurs exactly for the $V_P$ range, where the CoFeB gyrofrequency drops.
We attribute the observed upshift of the gyrotropic frequency in nonmagnetostrictive Py vortex to the strong dipolar coupling to the neighbouring CoFeB vortex through the Cu spacer. Essentially, this effect indicates that the dynamics of the Py vortex is pulled by the CoFeB vortex core gyration, as the strain-induced downshifted CoFeB resonance frequency approaches that of Py, upon reaching a certain threshold strain value. For $t_{Cu}$ = 10~nm, the CoFeB frequency downshift sets in for $V_P$ = 3~V, so does the Py frequency upshift [see Fig.~\ref{fig4}(d)].
Further increase of the voltage $V_P$ (and the generated piezostrain) to 5~V and above leads to the dynamical regime where both CoFeB and Py vortices gyrate at almost the same frequency demonstrating a synchronized dynamics as predicted by micromagnetic simulations~\cite{iurchukStressinducedModificationGyration2021}.

For the increased thickness of the Cu spacer, the threshold voltage for the CoFeB frequency downshift gradually increases [$\sim$6~V for $t_{Cu}$ = 20~nm and $\sim$8~V for $t_{Cu}$ = 40~nm, see Fig.~\ref{fig4}(e,f)]. For $t_{Cu}$ = 20~nm, the synchronized dynamical regime is observed for $V_P$ $\geq$9~V, whereas for $t_{Cu}$ = 40~nm, this regime was not reached at all, for the $V_P$ range covered in the dynamic experiments, due to weak dipolar coupling between CoFeB and Py layers.
The corresponding Py frequency upshift due to frequency pulling by the CoFeB layer, still visible for $t_{Cu}$ = 20~nm, vanishes for $t_{Cu}$ = 40~nm. This observation is in agreement with the measurements of Fig.~\ref{fig3}(d), where for $t_{Cu}$ = 40~nm, the CoFeB and Py vortices behave dynamically as noninteracting vorticies.

In conclusion, we demonstrated piezostrain-induced frequency pulling in stacked CoFeB/Py double-vortex structures,  magnetostatically coupled through the Cu spacer. We show, that the gyrotropic frequency of the nonmangetostrictive Py vortex can be efficiently pulled by the dynamics in the adjacent magnetostrictive CoFeB vortex, when two general conditions are met. First, the structure has to be designed so that the gyrotropic frequencies of the magnetostrictive and nonmagnetostrictive components are spaced closely enough to allow for the magnetoelastically downshifted gyrofrequency in the magnetostrictive vortex cross the gyrofrequency of the nonmagnetostrictive vortex. Second, the dipolar coupling between the two stacked vortices has to be sufficiently strong to allow for the strain-induced frequency pulling of the gyrofrequency of the nonmagnetostrictive vortex towards that of the magnetostrictive counterpart.

This result offers an additional degree of freedom for the control over dynamical regimes and synchronization conditions in spintronic oscillators, controlled by voltage and tunable by strain. 

This study is funded by the Deutsche Forschungsgemeinschaft (DFG, German Research Foundation) within the grant IU 5/2-1 (STUNNER) – project number 501377640. Support from the Nanofabrication Facilities Rossendorf (NanoFaRo) at the IBC is gratefully acknowledged. We thank Thomas Naumann for help with the sputtering of the CoFeB/Cu/Py stacks. 

\bibliographystyle{apsrev4-2}
\bibliography{references.bib}

\end{document}